\documentclass{article}
\usepackage{hyperref}
\usepackage{natbib}
\setcitestyle{numbers}
\setcitestyle{square}
\usepackage{amsmath}
\usepackage{amssymb}
\usepackage{amsthm}
\usepackage{mathtools}
\usepackage{braket}
\usepackage{bm}
\usepackage{bbm}
\usepackage{graphicx}
\usepackage{xcolor}
\DeclarePairedDelimiter{\abs}{\lvert}{\rvert}
\newcommand{\numberset}{\mathbb}

\newcommand{\R}{\numberset{R}}
\newcommand{\Z}{\numberset{Z}}
\newcommand{\C}{\numberset{C}}

\newcommand{\sigv}{\sigma_\text{v}}

\newcommand{\hsigd}{{\sigma}_\text{d}}

\newcommand{\bx}{\mathbf{x}}
\newcommand{\ri}{i}
\newcommand{\bC}{\mathbf{C}}
\newcommand{\bCo}{\mathbf{\bar{C}}}

\newcommand{\bF}{\mathbf{F}}

\newcommand{\bG}{\mathbf{G}}
\newcommand{\bH}{\mathbf{H}}

\newcommand{\bX}{\mathbf{X}}

\newcommand{\bna}{\boldsymbol{\nabla}}

\title{\large \bf Ericksen-Landau Modular Strain Energies \\ for Reconstructive Phase Transformations  in 2D crystals
}
\date{\small \today}
\author{{\normalsize Edoardo Arbib$^1$, Paolo Biscari$^1$},\\ {\normalsize Clara Patriarca$^2$, Giovanni Zanzotto$^3$}\thanks{email: \href{mailto:edoardo.arbib@mail.polimi.it}{edoardo.arbib@mail.polimi.it}, \href{mailto:paolo.biscari@polimi.it}{paolo.biscari@polimi.it}, \href{mailto:clara.patriarca@polito.it}{clara.patriarca@polimi.it}, \href{mailto:giovanni.zanzotto@unipd.it}{giovanni.zanzotto@unipd.it}}\\ \ \\ \small $^1$ Department of Physics, Politecnico di Milano, Italy\\ \small
$^2$  Department of Mathematical Sciences, Politecnico di Torino, Italy\\
\small $^3$ DPG, Universit\`a di Padova, Italy
}

\begin{document}

\maketitle

\noindent \null \hfill {\it To Jerry Ericksen,}\\
\noindent \null \hfill {\it always a precursor}

\begin{abstract}

By using modular functions on the upper complex half-plane, we study a class of strain energies for crystalline materials whose global invariance originates from the full symmetry group of the underlying lattice.
This follows Ericksen's suggestion which aimed at extending the Landau-type theories to encompass the behavior of crystals undergoing structural phase transformation, with twinning, microstructure formation, and possibly associated plasticity effects. Here we investigate such Ericksen-Landau strain energies for the modelling of reconstructive transformations, focusing on the prototypical case of the square-hexagonal phase change in 2D crystals. We study the bifurcation and valley-floor network of these potentials, and use one in the simulation of a quasi-static shearing test. We observe typical effects associated with the micro-mechanics of phase transformation in crystals, in particular, the bursty progression of the structural phase change, characterized by intermittent stress-relaxation through microstructure formation, mediated, in this reconstructive case, by defect nucleation and movement in the lattice.
\end{abstract}

\noindent \small \textbf{Keywords. }Reconstructive phase transformations, square-hexagonal transformation, crystal plasticity, Poincar\'e half-plane, Dedekind tessellation, Klein invariant, modular forms, deformation pathways

\section{Introduction}

Ericksen's early proposal \citep{eri77, eri80} of an infinite and discrete invariance group for a crystalline material's strain energy aimed at expanding Landau-type variational approaches to encompass structural phase transformations and twinning in crystals, with the associated phenomena of fine microstructure, and possibly defects, forming in the lattice. Accordingly, the material invariance of the crystalline substance should reflect the global symmetry of the underlying lattice, with the strain energy  density $\sigma$ invariant under all the deformations mapping the lattice onto itself. See also \cite{folkins} for a similar viewpoint. This invariance dictates the location of countably-many ground states for the crystal in strain space, including those produced by the lattice-invariant shears and rotations which play a key role in twinning mechanisms when in the presence of structural phase changes, as well as in lattice-defect creation and ensuing plastification phenomena \citep{bhattabook, PZbook, ContiZanzotto, BCZZnature, pacoreview,  pacoreview2, pacoreview3,  jamesreview}.

A wide-ranging extension of non-linear elasticity theory originated in this way, with special attention initially given to suitable ranges of finite but not too-large deformations, i.e., to `Ericksen-Pitteri neighborhoods' (EPNs) in strain space \citep{eri80, pitterireconciliation, PZbook, ContiZanzotto, bhattabook}, whereon the global lattice invariance reduces to point-group symmetry.\footnote{\label{definition} These domains were considered in \cite{eri80} to reconcile the present approach to crystal mechanics with the Laudau-type theories based on standard point-group invariance \citep{pitterireconciliation, PZbook, ContiZanzotto}. Structural phase transitions are termed 'weak' when their spontaneous transformation strains are confined to suitable EPNs. Finite deformations within these domains cause symmetry breaking in the distorted lattices, and the parent and product lattices' point-group symmetries are in a group-subgroup relation. When this does not happen the phase change is reconstructive. Most relevant examples of the latter are the bcc-fcc or bcc-hcp transformations in 3D \citep{toledanoreconstructive},
and the $s$-$h$ transformation in 2D Bravais lattices, see \cite{ContiZanzotto} for more details.}
A large body of literature originated from such EPN-based approach, especially aiming at modelling reversible martensitic transformations \citep{balljames2, bhattabook, PZbook, luskin, dolzmannbook, jamesperspective1, jamesreview}, also with the goal of improving the mechanical properties of shape-memory alloys, for instance to enhance their reversibility performance through the control of twinned-microstructure formation \citep{13james_nat, jamesperspective2}.

Another line of research used the above Ericksen-Landau framework to model a wider class of phenomena in crystal mechanics, including reconstructive structural transformations where strains may attain or go beyond the EPN bundaries, producing defect nucleation and evolution in the lattice, and, in general, also to model phenomena directly related to the plastic behavior of crystalline materials, where the large deformations are not confined to any EPNs in strain space. Although discussions of the behavior of 3D crystals based on global lattice symmetry can be found in \citep{GZborn, BCZZnature, acklandniti, ironshear2, biurzaza}, more systematic research has been done been done on crystal elasto-plasticity only in the 2D case. A family of Ericksen-Landau energies 2D crystals, obtained by patching suitable polynomials to obtain $\mathcal{C}^{2}$-smoothness and global invariance, was proposed in \citep{ContiZanzotto}. This allowed an improved understanding of the behavior of crystalline materials also in regimes of large deformations possibly outside the EPNs   \citep{pacoreview, pacoreview2, pacoreview3, pacoreview4, PRLgruppone, comptrend}.

A parallel line of work on the 2D case was based on the observations in \cite{parry, folkins}, where the natural tools proposed for a theory encompassing full lattice symmetry in 2D are modular functions, the well known class of complex maps arising in diverse branches of Mathematics and Physics \citep{modularforms2, modularforms1, dedekind3, dedekind1, modularphysics}. This led to the formulation of potentials suitable for 2D crystal plasticity, by following, in particular, the suggestion in \citep{parry} to construct Ericksen-Landau energies by means of the 'Klein modular invariant' $J$ \citep{modularforms2, modularforms1, mathematicaJ, wikiJ}. This is akin to using a modular order parameter for crystal mechanics, extending earlier related notions such as the 'transcendental order parameter' in \cite{transcendental1, transcendental2}. In this spirit, \cite{PRLgruppone, IJP} consider $J$-based strain energies with a unique ground state, up to full lattice symmetry, exploring the ensuing variational modelling of 2D crystal elasto-plasticity.

Here we continue these investigations by examining an explicit, simplest class of Ericksen-Landau $J$-based strain potentials for reconstructive  transformations in 2D lattices, focussing on the most relevant case of energy functions exhibiting ground states with the two maximal symmetries, square and hexagonal ($s$-$h$), of 2D Bravais lattices. We explore some basic properties of the $s$-$h$ strain-energy landscapes moulded by global symmetry, in particular their bifurcation and valley-floor network, which are important in the selection of the activated deformation paths under total-energy minimization. We thus use a global $s$-$h$ potential in the simulation of a quasi-static shearing test, obtaining typical effects associated with the micromechanics of phase transformations in crystals. In particular, we observe strain avalanching underpinned by bursty coordinated basin-hopping activity of the local strain values under the slowly changing boundary conditions. This produces the inhomogeneous progression of the structural phase change, characterized by jagged stress relaxation via bursty microstructure development in the body, also mediated, in the reconstructive case, by defect nucleation and movement in the lattice. The present simulations also confirm the role, highlighted yet in \citep{IJP}, of the energy's  valley floors as largely establishing the deformation pathways for a crystal's intermittent evolution under an external driving.

\section{Strain energies for 2D crystalline materials}

\subsection{The strain energy of 2D crystals \\ and Ericksen's proposal for its invariance}

We consider a two-dimensional (2D) hyperelastic material, whose deformations are one-to-one maps $\bx = \bx(\bX)$, where the Cartesian coordinates $(x_1,x_2)$ identifying the current positions of material points $\bX=(X_1,X_2)$ in a given reference state are considered with respect to a given ortho-normal basis $\{\bold u_1,\,\bold u_2\}$. The deformation gradient $\bF=\bna\bx$ has matrix elements $F_{ij}=\partial x_i/\partial X_j$, and $\bC=\bF^T\bF = \bC^T>0$ is the symmetric, positive-definite the Cauchy-Green strain tensor. The strain-energy density $\sigma$ is a smooth real function of $\bC$, and satisfying the material-symmetry requirements \eqref{eq:energyinvariance}$_1$-\eqref{eq:energyinvariance}$_3$:
\begin{equation}
\sigma = \sigma(\bC)=\sigma(\bG^T\bC\bG), \quad \bG \in {\cal G}, \quad {\cal G} = E^{-1}GL(2,Z)E,
\label{eq:energyinvariance}
\end{equation}
to hold for any $\bC$ and for any tensor $\bG$ in a suitable group $\cal G$ characterizing the response of the material. For crystalline substances we assume with Ericksen that the invariance group $\cal G$ be dictated by the material's underlying lattice structure \citep{eri77, eri80}, see also \cite{folkins, michel, PZbook, jamesreview}. In the 2D case under consideration here, this means that $\cal G$ should be a suitable conjugate to the group describing the global symmetry of 2D Bravais lattices, as is made explicit in Eq. \eqref{eq:energyinvariance}$_4$ above, with $E = (e^h_j)$, for $\bold e_j = e^h_j \bold  u_h$ (summation understood, with $j,h$ = 1, 2), where $\{\bold e_1,\,\bold e_2\}$ are the lattice basis in the reference state, and $\text{GL}(2,\Z)$ denotes the group of unimodular (thus invertible) 2 by 2 matrices with integral entries \citep{PZbook}. For brevity we refer to assumption (\ref{eq:energyinvariance})$_4$ as to the GL-invariance of the density $\sigma$ in (\ref{eq:energyinvariance}), which we split into the sum of a convex volumetric part $\sigv$, penalizing the departure of $\det\bC$ from 1, and a distortive term $\hsigd$ depending on the unimodular tensor $\bCo=(\det\bC)^{-1/2}\bC$:
\begin{equation}
 \sigma(\bC)=\sigv(\det\bC)+\hsigd(\bCo).
\label{eq:sigelastic}
\end{equation}
Due to the GL-periodicity (\ref{eq:energyinvariance})$_4$, $\hsigd$ in (\ref{eq:sigelastic}) is non-convex and only needs to be defined on a GL-fundamental domain in the space of unimodular strains, such as $\mathcal{D}$ made explicit in (\ref{fundomain}) below.

\subsection{Modular forms and GL-invariant strain energies \\ on the  Poincar\'e half plane}

\begin{figure}
\centering
\includegraphics[height=4.8cm, width=8.5cm]{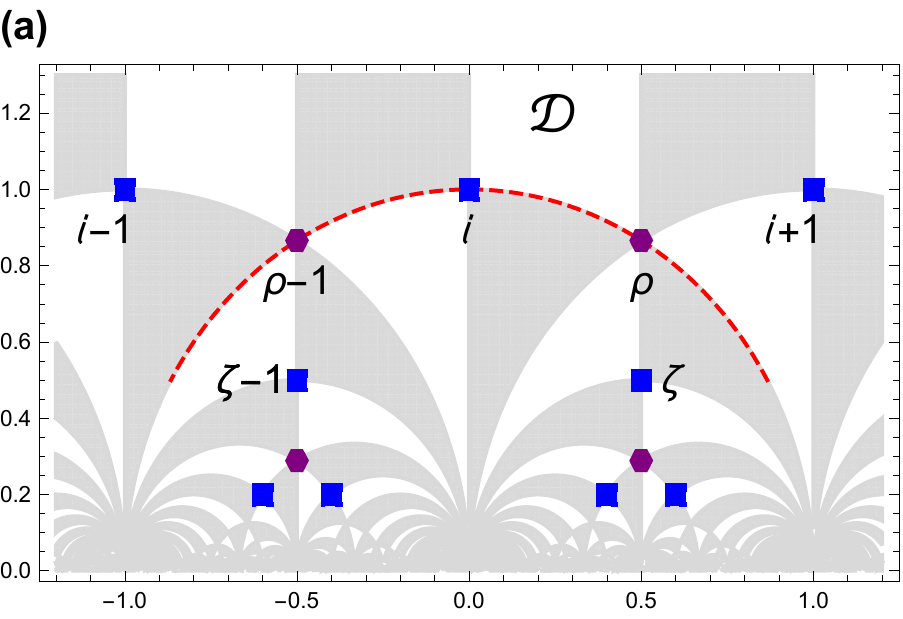}
\quad
\includegraphics[height=4.8cm, width=10cm]{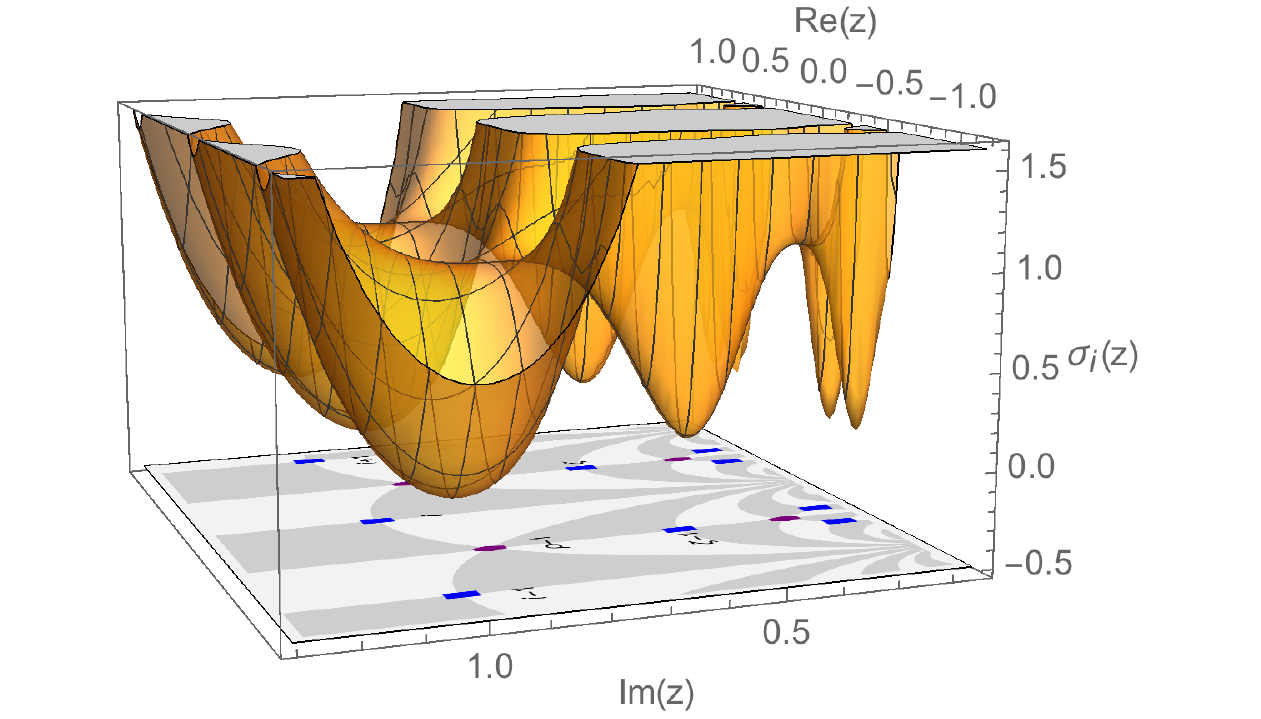}
\caption{\small (Color online) (a) Dedekind tessellation of the Poincar\'e half-plane $\mathbb{H}$. Gray or white domains represent GL-copies of the fundamental domain $\cal D$ in \eqref{fundomain}. The positions are indicated of nine GL-equivalent square points \big($i$, $i + 1$, $\zeta= \frac{1}{2}(i+1)$, $\zeta+1$, ..., in blue\big),
and four GL-equivalent hexagonal points \big($\rho = e^{i \pi/3}$, $\rho -1$, ..., in purple\big). The GL-equivalent points appear to be getting closer to each other the closer they get to the real axis, but are equidistant in the hyperbolic metric (\ref{hyperbolicmetric})$_2$ on the strain space $\mathbb{H}$. This rectangle and the indicated red dashed arc of the unit circle about the origin of $\mathbb{H}$ refer also to~Fig.~\ref{fig:recons}.
(b) Plot of the GL-periodic strain energy with square minimizers, see the function $\sigma_{i}$ in \eqref{eq:sq} with $\mu=1$. The plot is on the rectangle in panel (a), so that we observe nine energy wells, with bottoms at the nine blue square GL-equivalent points of panel (a). }
\label{fig:Dedekind}
\end{figure}

Due to their GL-invariance, smooth potentials as in (\ref{eq:energyinvariance})-(\ref{eq:sigelastic}) are closely related to the modular functions on the Poincar\'e upper complex half-plane $\mathbb{H}$ \citep{folkins,parry}. This is best seen by smoothly mapping the space of 2D unimodular (positive-definite, symmetric) strain tensors $\bCo$ bijectively to $\mathbb{H}$:
\begin{equation}
\hat z(\bCo)={\bar{C}}_{11}^{-1}({\bar{C}}_{12} + i) \in \mathbb{H},
\label{bijectionH}
\end{equation}
\begin{equation}
\mathbb{H} = \{ x+\ri y\in\C, y>0 \},
\qquad (ds)^2=\big[(dx)^2+(dy)^2\big]/y^2 \;,
\label{hyperbolicmetric}
\end{equation}
where (\ref{hyperbolicmetric})$_2$ is the standard 2D hyperbolic metric on $\mathbb{H}$ \citep{poincarehalfplane2, poincarehalfplane1, poincarehalfplane3}, and $C_{ij}$ are the components of $\bC$ in the basis $\{\bold u_1,\, \bold u_2\}$. In the above complex parameterization of strain space the material-symmetry maps $\bC \mapsto \bG^T\bC\bG$, for $\bG \in {\cal G}$, correspond to the action on $\mathbb{H}$ of the isometries of $\mathbb{H}$ (linear fractional transformations) with integral entries, supplemented by the map $z \mapsto -\bar z$, see \cite{IJP} for details. The Dedekind tessellation of $\mathbb{H}$ shown in Fig.~\ref{fig:Dedekind} \citep{dedekind1, dedekind5} represents geometrically this action, evidencing the $\text{GL}(2,\Z)$-related congruent copies of the fundamental domain\footnote{The structure of $\mathcal{D}$ summarizes the (unimodular) strains giving all the possible ways in which a 2D Bravais lattice can be deformed, up to GL-symmetry. The interior of $\mathcal{D}$ corresponds to strain tensors producing lattices with trivial (oblique) symmetry; points on the boundary $\partial\mathcal{D}$ are associated with strains generating lattices with nontrivial symmetries, including rectangular and rhombic lattices; finally the corner points $i$ and $\rho= e^{i \pi /3}$ correspond to strains giving respectively a square and a hexagonal lattice. See \cite{parry}, \cite{ContiZanzotto} for further details, also on the relation of the fundamental domain \eqref{fundomain} to the EPNs mentioned in the Introduction.}
\begin{equation}
\label{fundomain}
\mathcal{D}=\{ z\in \mathbb{H} : \abs{z}\ge 1,\ 0\le\text{Re}(z)\le\tfrac{1}{2} \}
\end{equation}
on $\mathbb{H}$. In light of this, \cite{parry} suggested that, as a main building block to obtain GL-invariant smooth potentials $\hsigd$ in (\ref{eq:energyinvariance}), one can use the a well-known 'Klein invariant' $J$, which is a $\text{SL}(2,\Z)$-periodic holomorphic function on $\mathbb{H}$ \citep{modularforms2, modularforms1,wikiJ,mathematicaJ}, the modular group $\text{SL}(2,\Z)$ being the positive-determinant subgroup of $\text{GL}(2,\Z)$. Indeed, owing to the properties of $J$, it is possible to consider  $J$-based Ericksen-Landau strain-energy functions $\hsigd$ as in  (\ref{eq:energyinvariance})-(\ref{eq:sigelastic}) by setting
\begin{equation}
\hsigd(\bCo) =\hsigd\big(J(\hat{z}(\bCo)) \big),
\label{Jbased}
\end{equation}
where the smooth function $\hsigd(J)$ should be such that it guarantees \citep{IJP}: (a) the full GL-periodicity (\ref{eq:energyinvariance}) for (\ref{Jbased}), rather than the sole invariance under $\text{SL}(2,\Z)$ exhibited by $J$; and, (b) the existence of a positive-definite elastic tensor for any stable lattice configuration. Examples of such potentials, suitable for the elasto-plasticity of 2D crystals, and for their reconstructive transformations, are discussed explicitly hereafter.

\subsection{Strain energies for 2D crystal plasticity}
\label{plasticityenergiesIJP}

In \cite{PRLgruppone, IJP} were analyzed some simplest forms of GL-invariant $J$-based strain-energy functions $\hsigd$ as in (\ref{Jbased}), exhibiting a single ground-state configuration in each GL-copy of the fundamental domain $\cal D$. This is a particularly relevant class of strain potentials, which can be used to model elasto-plastic phenomena in crystalline materials.

Let the (unique, up to $\cal G$-symmetry) equilibrium configuration for the lattice be given by the strain $\bCo_0$, and set $z_0 = \hat{z}(\bCo_0) \in \cal D$. Then for any $z_0$ such that $J'(z_0)\neq 0$ (that is, for any ground state except for square and hexagonal ones: $z_0 \neq i, \rho$) the simplest $J$-based GL-invariant strain-energy function in (\ref{Jbased}) has the form:
\begin{equation}
\sigma_{z_0}\big(\bCo\big) = \mu |J(z) - J(z_0)|^2,
\label{eq:nonsing}
\end{equation}
where $\mu >0$ is an elastic modulus, and $z = \hat{z}(\bCo)$ as in (\ref{bijectionH}). In the case of the maximally symmetric ground states $z_0=i$ (square) or $z_0=\rho$ (hexagonal), taking into account that $J(i) = 1$, $J(\rho)= 0$, $J'(i) = 0$, $J'(\rho) =J''(\rho) = 0$, we have that the simplest $J$-based GL-potentials with non-degenerate elastic moduli at their minimizers are respectively given by:
\begin{align}
&\sigma_{i}\big(\bCo\big)= \mu |J(z)-1| &&\text{for }\; z_0=i \;\; \text{(square)}
\label{eq:sq}\\
&\sigma_{\rho}\big(\bCo\big)= \mu |J(z)|^{2/3} &&\text{for }\; z_0=\rho \;\; \text{(hexagonal)}.
\label{eq:hex}
\end{align}
As an example, we show in Fig.~\ref{fig:Dedekind} a portion of the GL-periodic energy landscape on $\mathbb{H}$ given by the square energy \eqref{eq:sq}.

\section{Ericksen-Landau theory for reconstructive \\ transformations in crystalline materials}
\label{section reconstructive}

\subsection{Simplest $J$-based strain energies for the \\ square-hexagonal transformation}

The functions in \eqref{eq:nonsing}-\eqref{eq:hex} can be used to construct GL-potentials of the type \eqref{Jbased} with more than one minimizer in the fundamental domain $\cal D$, so that they are suitable for crystals which may undergo structural phase changes between different stable equilibrium configurations of their lattice (see Footnote (3)). *****Here we consider explicitly the case of reconstructive transformations, and their most relevant case, in which the two energy minimizers in $\cal D$ are the maximally symmetric points $i$ (square) and $\rho$ (hexagonal). This will produce a GL-invariant potential suitable for the $s$-$h$ transformation in 2D Bravais lattices.\footnote{Strain energies for weak (symmetry-breaking) martensitic transformations (see Footnote (1)) involve the presence of several variant wells inside the EPNs \cite{PZbook, bhattabook, jamesreview}, and thus have more complex forms than the 2D potentials considered hereafter in \eqref{eq:rec}, and which can still be written in terms of the functions in \eqref{eq:nonsing}-\eqref{eq:hex}, as is preliminarily discussed in \cite{tesipatriarca}.}

\begin{figure}
\centering
\includegraphics[height=5.5cm, width=11cm]{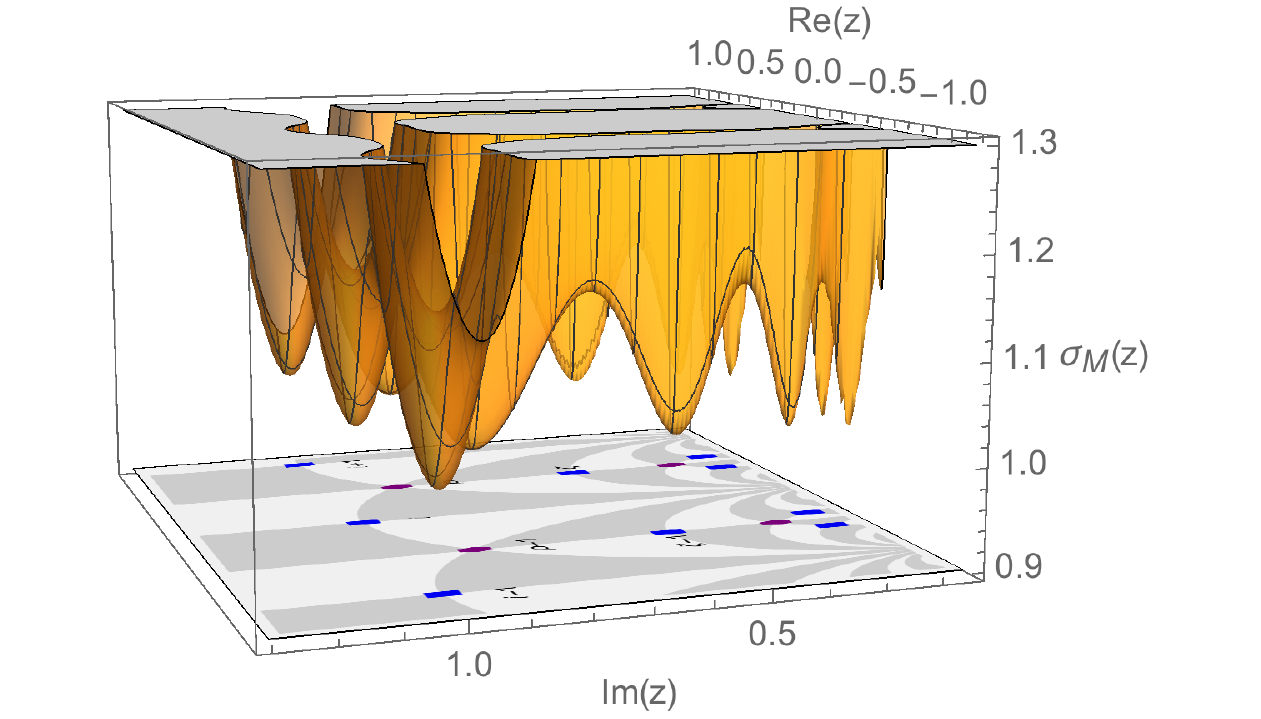}
\quad
\includegraphics[height=4cm]{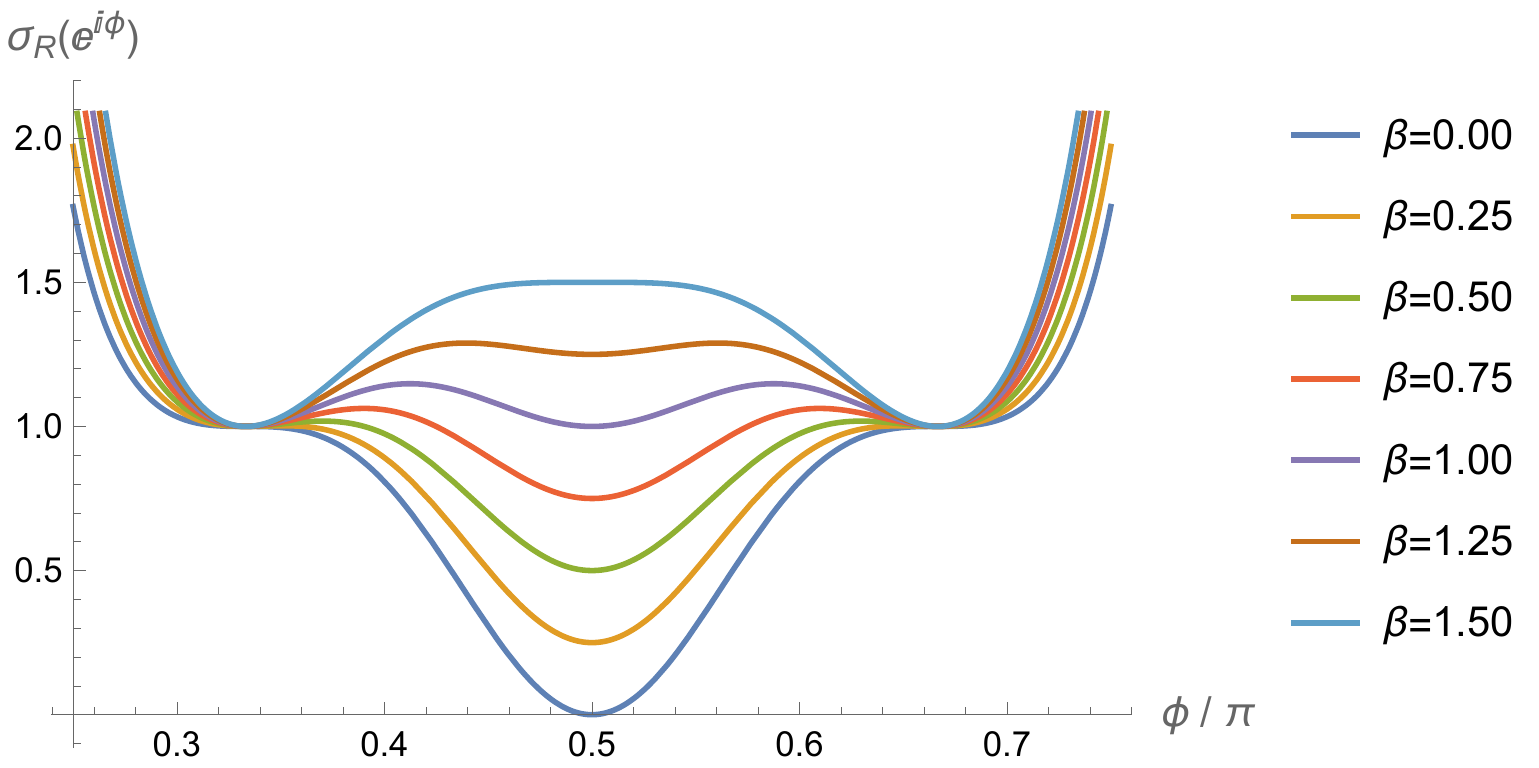}

\caption{\small (Color online) GL-invariant strain energy for the square-hexagonal phase transformation. (a)  Plot of the $s$-$h$ potential $\sigma_\text{R}$ in Eq. \eqref{eq:rec} for $\beta= 1$ and $\mu=1$, on the domain in  Fig.~\ref{fig:Dedekind}(a), whereon the energy $\sigma_\text{R}$ exhibits thirteen energy wells: nine well bottoms are located at the nine equivalent square points, and four at the equivalent hexagonal points, shown in Fig.~\ref{fig:Dedekind}(a).
(b) Section of the plot of the $s$-$h$ energy $\sigma_\text{R}$ in \eqref{eq:rec} with $\mu=1$ and varying $\beta$, taken along the unit-circle shown in red in Fig.~\ref{fig:Dedekind}(a).
The value $\phi=\frac{\pi}{2}$ corresponds to the square point $i$, while $\phi=\frac{\pi}{3}$ and $\phi=\frac{2\pi}{3}$ correspond to the two neighbouring hexagonal points $\rho-1$ and $\rho$, with rhombic points given by generic values of $\phi$, see also Fig.~1 and Footnote (2).}
\label{fig:recons}
\end{figure}

A simplest class of such $s$-$h$ densities is given by the following normalized linear combination of the two functions in \eqref{eq:sq}-\eqref{eq:hex}:
\begin{equation}
\begin{split}
    \sigma_\text{R}\big(\bCo\big)&=\sigma_{i}\big(\bCo\big)+\beta\sigma_{\rho}\big(\bCo\big)\\
    &=\mu |J(z)-1|+\beta\mu |J(z)|^{2/3}  ,
    \label{eq:rec}
\end{split}
\end{equation}
where $z=\hat{z}(\bCo)$ as in (\ref{bijectionH}) and \eqref{Jbased}, and where we consider $\beta>-\frac32$. The modulus $\mu$ is here a scale factor which will henceforth be set to 1, so that \eqref{eq:rec} defines a one-parameter family of potentials with critical points $i$ and $\rho$, whose relative height is controlled by $\beta$ as shown in Fig.~\ref{fig:recons}.

\subsection{Bifurcation and valley floors}

\begin{figure}
\centering
\includegraphics[height=4.1cm, width=6cm]{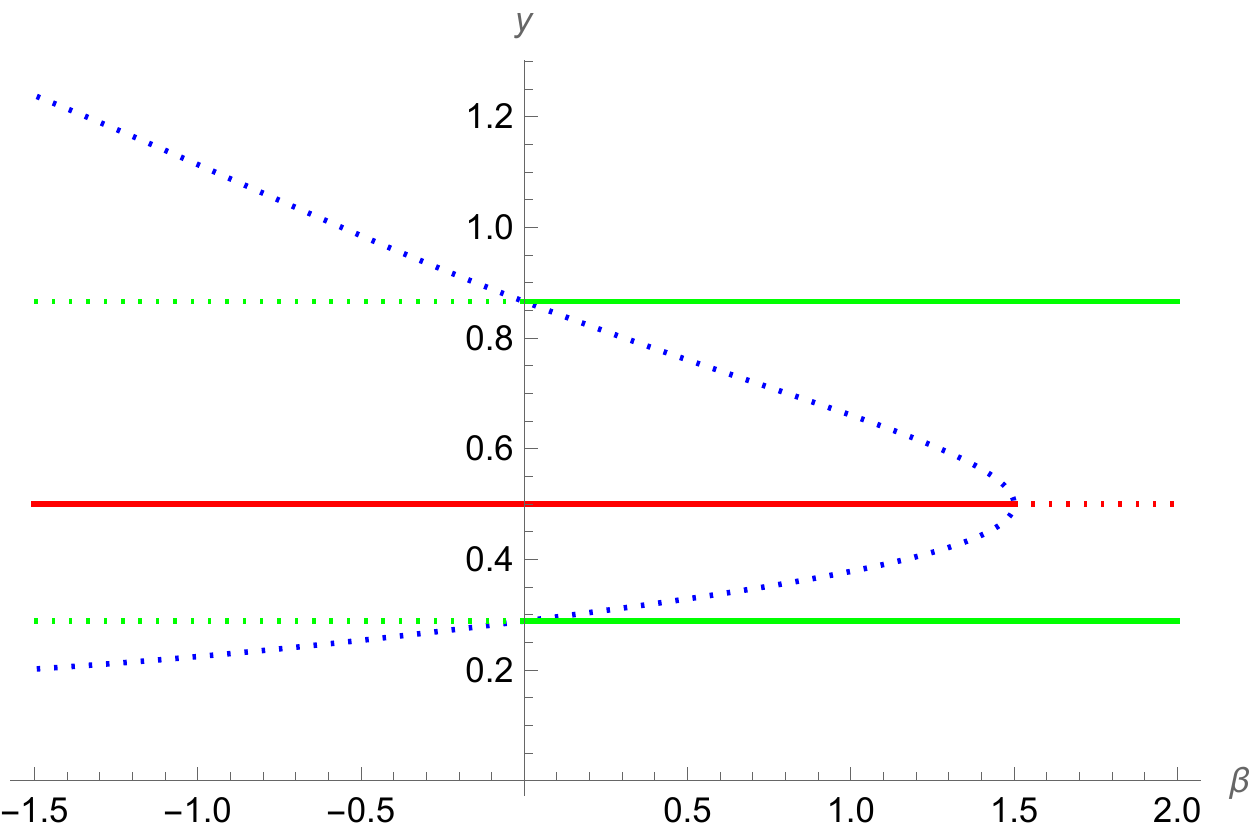}
\includegraphics[width=5.4cm]{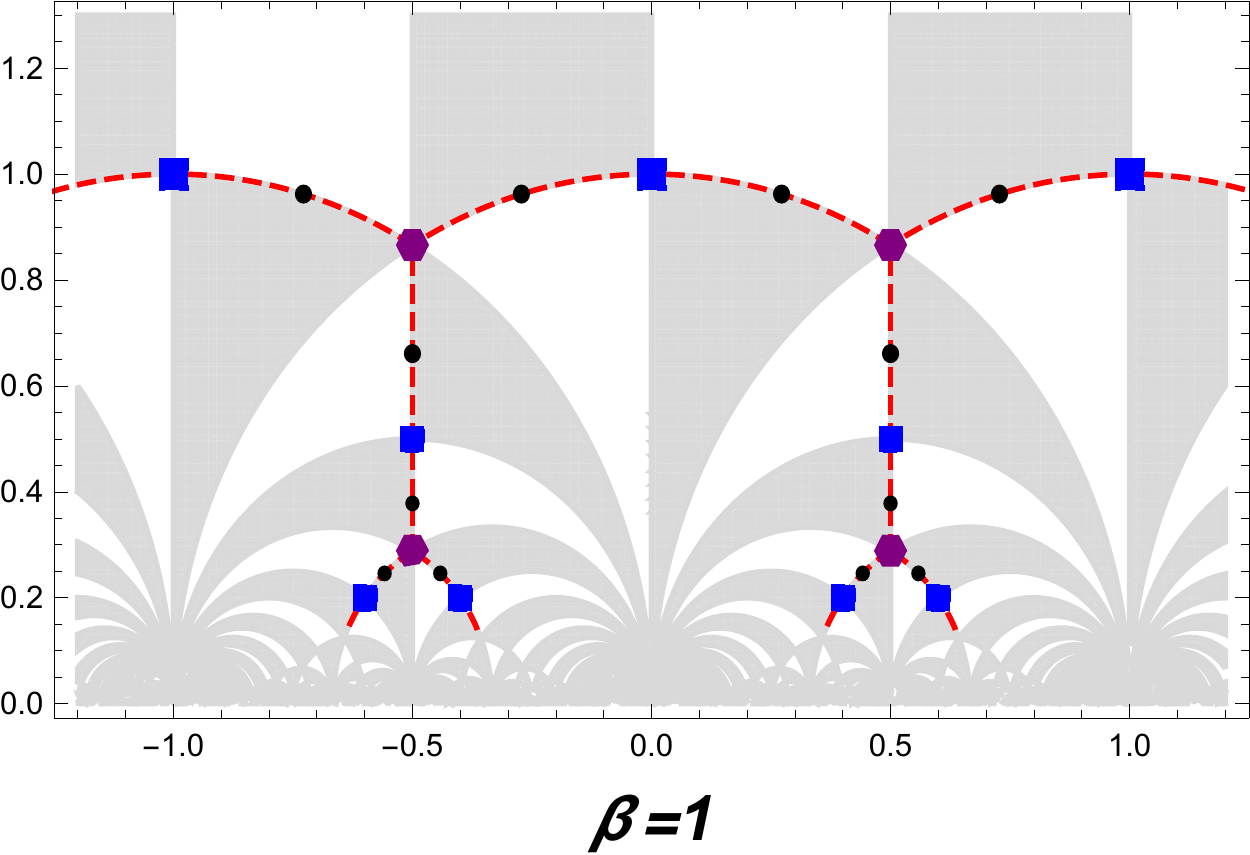}

\caption{\small (Color online) Bifurcation and valley floors for the $s$-$h$ energies. (a) Section on the plane ($\beta$, $y$), for $x=\frac12$, of the GL-invariant bifurcation diagram for the critical points of the $s$-$h$ energy  $\sigma_\text{R}$
in Eq.~\eqref{eq:rec} for $\beta > -\frac32$. Dotted and solid lines indicate unstable and stable critical-point branches, respectively,
with the square-$i$ (solid red line) and hexagonal-$\rho$ (solid green lines) which coexist as local minimizers for $0<\beta<\frac32$, with rhombic saddles in between (dotted blue lines). See also Video V1 in the SM. (b) GL-invariant hyperbolic network of the valley floors on $\mathbb{H}$ for the strain energy $\sigma_{\text{R}}$ in Eq. (\ref{eq:rec}), for $\beta =1$. Nodes (blue and purple symbols) are at the $s$-$h$ minimizers, and edges are along those geodesics on $\mathbb{H}$ which contain a pair of $s$-$h$ minimizers (for clarity only the arcs joining such $s$-$h$ points are marked, by dotted red lines). The rhombic saddles mentioned in panel (a) are marked by black dots. See also Figs.~\ref{fig:snapshots}-\ref{nubecanyon}, and Videos V2, V3 \cite{video2, video3} in the SM.
}
\label{bifurcationvalleyfloors}
\end{figure}

We show in Fig.~\ref{bifurcationvalleyfloors}(a) the bifurcation on the plane ($\beta$, $y$), with $x=\frac12$, for the critical points of the $s$-$h$ energy $\sigma_\text{R}$ in \eqref{eq:rec}. This is obtained from the way the global GL-symmetry of the potential constrains, via the implied local (point-group) symmetry, the second-derivatives of its critical and bifurcation points \citep{eri80, PZbook}. This diagram expectedly has the same main features as the one pertaining to the polynomial-based $s$-$h$ energy in \cite{ContiZanzotto}. The actual GL-periodicity of the bifurcation pattern of the energy in \eqref{eq:rec} is sketched in Video V1 of the Supplementary Material (SM), see \cite{video1}.

In Fig.~\ref{bifurcationvalleyfloors}(a) we see that the square critical point $i$ of $\sigma_\text{R}$ is stable for $\beta<\frac32$, losing stability at $\beta = \frac32$ through a subcritical pitchfork to two rhombic critical points (saddles). On the other hand, $\rho$ is a minimum for $\beta>0$, becoming unstable at $\beta=0$, where symmetry dictates the presence of a monkey saddle, unfolding \citep{electronsmonkeysaddle} via a transverse bifurcation to three rhombic critical-point branches (standard saddles, only one of which belongs to the plane ($\beta$, $y$) of the figure; see also Video V1 in the SM). The points $i$ and $\rho$ are the only local minimizers of $\sigma_\text{R}$ in $\cal D$ for $0<\beta<\frac32$, and in this range the energy $\sigma_\text{R}$ in \eqref{eq:rec} is thus suitable to model the $s$-$h$ transformation. The coexisting minima $i$ and $\rho$ have the same energy at the Maxwell value $\beta = \beta_M = 1$, so that the global minimum is $i$ for $0\leq\beta\leq1$, while it is $\rho$ for $1\leq\beta\leq\frac32$.

For the $s$-$h$ energy $\sigma_\text{R}$ in \eqref{eq:rec} with $0<\beta<\frac32$, it is also interesting to highlight the structure of the infinite GL-invariant network of the valley floors connecting the energy extremals on $\mathbb{H}$, which, as mentioned in the Introduction, considerably inform us regarding the evolution of the strain field in $\mathbb{H}$. Precisely, the valley floors of an energy $\sigma$ are the gradient-extremal loci on $\mathbb{H}$ locally satisfying $\bH\,\bna\sigma  - \xi\, \bna\sigma  =0$, with $\xi \in \R$ and $\bH\bold v\cdot\bold v\geq 0$ \citep{valleyfloor1, valleyfloor2, valleyfloor3}, where $\bna\sigma$ and $\bold H$ are respectively the gradient and Hessian of $\sigma$, and $\bold v$ is any vector orthogonal to the energy gradient, with derivatives and orthogonality considered in relation to the hyperbolic metric (\ref{hyperbolicmetric})$_2$. This produces the following explicit gradient-extremal equation:
\begin{equation}
\frac{2}{y^3}\,|\bna\sigma|^2\,     \bold v_2  -\frac{2}{y^2}
\bH\,\bna\sigma  + \xi\, \bna\sigma  =0,
\label{explicit valleyfloors}
\end{equation}
where $\bold v_2$ is the unit vector in the $y$-direction on $\mathbb{H}$, and derivatives are intended in its standard atlas $(x, y)$. For $\beta=1$ the valley floors of the $s$-$h$ energy $\sigma_\text{R}$ obtained from \eqref{explicit valleyfloors} compose the hyperbolic network highlighted in Fig.~\ref{bifurcationvalleyfloors}(b), whose edges lie on those geodesics of $\mathbb{H}$ \citep{poincarehalfplane2, poincarehalfplane1} which contain both the $s$-$h$ minimizers.

\section{Shear-driven $s$-$h$ transformation}

We investigate numerically the behavior of an $s$-$h$ phase-transforming crystal in quasi-static shearing, imposed to the top side of a square body containing a coaxial square lattice, with fixed bottom side and the remaining two sides free. In this incremental test, for each value of the shear parameter $\gamma$ a local minimizer of the body's total strain-energy functional is computed through the density $\sigma_\text{R}$ in \eqref{eq:rec} with $\beta =1$, and complying with the imposed boundary conditions (see \cite{IJP}  for computational details).

\begin{figure}
\centering
\includegraphics[width=10.75cm]{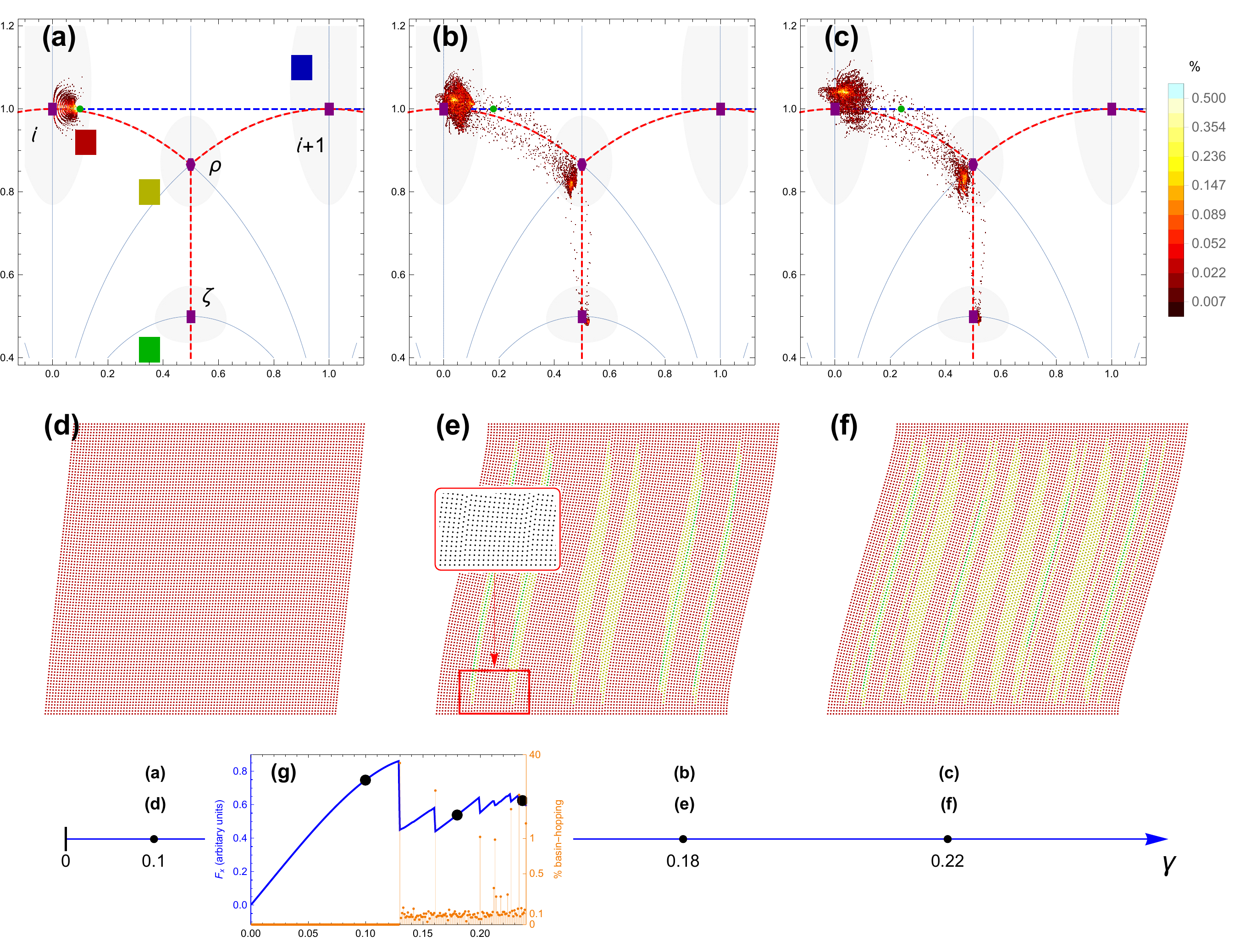}
\caption{\small (Color online) Shearing of an $s$-$h$ phase-transforming crystal, with strain energy $\sigma_\text{R}$ in \eqref{eq:rec} for $\beta =1$. The imposed loading is along a primary-shear direction in the square lattice, parallel to the driven horizontal body-sides. The associated path in $\mathbb{H}$ is the straight dashed blue line $i \rightarrow 1+i$ in panels (a),~(b),~(c), with increasing shear parameter $\gamma$ (green dot), from the defect-free reference configuration in the ground state $z_0=i$ ($\gamma =0$) to the GL-equivalent fully-sheared square configuration $i+1$ ($\gamma =1$). Convexity domains around each $s$-$h$ energy minimizer are shaded gray, and the valley floors of $\sigma_\text{R}$ are in dashed-red as in Fig.~\ref{bifurcationvalleyfloors}(b). The snapshots (a),~(b),~(c) show the evolution of the strain clustering during shear, given by the heatmap 2D-histogram for the cell-strain density evolving on the Dedeking tessellation of $\mathbb{H}$ in Fig.~\ref{fig:Dedekind}(a). Panels (d),~(e),~(f) show the associated body deformation (whreon the color coding for the cell strains in $\mathbb{H}$ is as in panel (a)), with panel (g) displaying the stress-strain relation (blue jagged line) for increasing $\gamma$. The response is elastic to about $\gamma = 0.13$, after which a bursty phase-transformation regime begins. Panels (d),~(e),~(f) show this is marked by the formation of $s$-$h$ phase mixtures (twin bands and lath-type microstructures) mediated by evolving lattice defects, as seen in the detail inset to panel (e). The deformation's intermittency is tracked in panel (g) through the sequence of relaxation events given by the jumps in the jagged stress-strain diagram, as well as the orange spikes indicating the percentage of cell-strain values that are hopping energy basin for each value of $\gamma$. Panels (a),~(b),~(c) show that the strain-cloud path on  $\mathbb{H}$ under the imposed boundary condition here follows the directions of the valley floors in the energy landscape, as is the case in crystal plasticity \citep{IJP}. More information on the bursty deformation triggered in this shear test is also in Figs. \ref{nubecanyon}-\ref{cdot} and Videos V2, V3, V4 in the SM \cite{video2, video3, video4}. }
\label{fig:snapshots}
\end{figure}

Figs.~\ref{fig:snapshots}(d)-(e)-(f) show three snapshots of the resulting $\gamma$-dependent strain field in the sheared crystal. Figs.~\ref{fig:snapshots}(a)-(b)-(c) highlight the corresponding strain clustering as a cloud of points which evolves with $\gamma$ on the GL-domains of the Dedekind tessellation of $\mathbb{H}$. See Video V2 \cite{video2} in the SM for the numerical simulation of shearing up to $\gamma = 0.24$.

The stress-strain relation in Fig.~\ref{fig:snapshots}(g) shows that the initially defect-free lattice begins shearing with a significant elastic charge, the associated strain cloud widening away from $i$b in $\mathbb{H}$, as $\gamma$ moves away from 0, due to the growing strain heterogeneity caused by the unloaded body-sides, see Figs.~\ref{fig:snapshots}(a)-(d). A large transformation event at about $\gamma = 0.13$ ends the elastic regime, with a large stress drop taking place as part of the strain cloud in $\mathbb{H}$ splits away from its initial location near the reference state $i$ towards the neighbouring well in $\rho$. A portion of the cells' strains remain far from the well bottoms, elastically stabilized on the intermediate non-convex regions, see Figs.~\ref{fig:snapshots}(b)-(c).

\begin{figure}
\centering
\includegraphics[width=8.0cm]{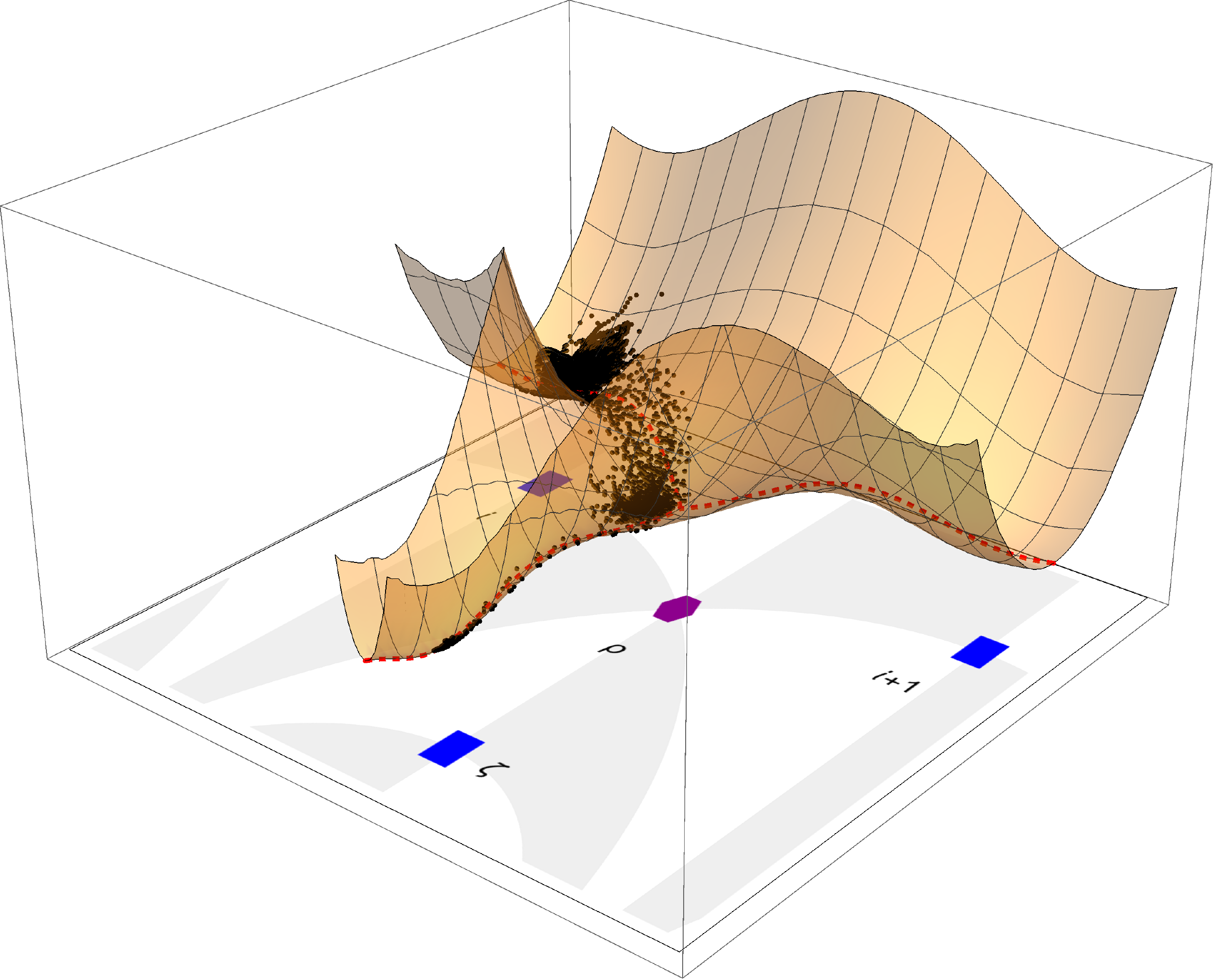}
\caption{\small
(Color online) Snapshot of the bursty evolution of the strain field on $\mathbb{H}$ ($\gamma = 0.24$) shown directly on a portion of the energy landscape during the shear-driven $s$-$h$ transformation. The GL-energy is $\sigma_\text{R}$ in Eq.~(\ref{eq:rec}) with $\beta = 1$, so the $s$-$h$ wells $i$, $\rho$, $\zeta$, $i+1$, ..., all have the same depth. The avalanching strain values largely follow the energy valley floors indicated in dashed red, as in Fig.~\ref{bifurcationvalleyfloors}(b). See also Video V3 \cite{video3} in the SM.
}
\label{nubecanyon}
\end{figure}

From there on, the imposed shearing induces a bursty deformation process in the body, characterized by an intermittent sequence of stress-relaxation events due to avalanching $s$-$h$ microstructure formation assisted by lattice-defect evolution, as can be seen in Figs.~\ref{fig:snapshots}(e)-(f)-(g), and Video V2 in the SM \cite{video2}.
These phenomena occur as the strain field in the lattice locally takes advantage, for the relative minimization of the total energy, of the available $s$-$h$ GL-wells of the density $\sigma_\text{R}$. Twin-type bands and dislocations emerge as neighboring lattice cells suitably stretch or shear and rotate while satisfying Hadamard's kinematic compatibility, together with the imposed boundary conditions.\footnote{The presence of defects in the lattice as a consequence of the phase change, as shown in Fig.~\ref{fig:snapshots}, leads to the irreversibility typically observed in reconstructive transformations, in both experiment and simulation \citep{BCZZnature, laguna1}.}
Long-range elastic interactions correspondingly produce coordinated basin-hopping on $\mathbb{H}$ which results in strain avalanches within the shearing body under the slow driving, see also \cite{PRLgruppone, IJP, comptrend}.
These complex deformation mechanisms involving both phase transition and defect evolution, occur here with no need for auxiliary hypotheses: they originate directly from energy minimization not only due to the GL-arrangement of the density minimizers in strain space, but also to the way in which the global GL-sym\-metry shapes the energy landscape. Indeed, we observe here in Figs.~\ref{fig:snapshots}(e)-(f)-(g), as is the case also with crystal plasticity \citep{IJP}, that the GL-energy valley floors act as strain-cloud deformation pathways on $\mathbb{H}$. For instance, the creation and evolution of the vertical shear bands in Figs.~\ref{fig:snapshots}(e)-(f), is due to strain avalanches occurring as suitable lattice domains leave the square reference state $i$, with the strain cloud following the valley-floor path $i \rightarrow \rho \rightarrow \zeta$ in $\mathbb{H}$, see Figs.~\ref{fig:snapshots}(b)-(c), although the body is being externally loaded in the horizontal principal shear direction $i \rightarrow i +1$ (green dot in the same figures) for the square crystal. The activation of the deformation pathway $i \rightarrow \rho \rightarrow \zeta$ implies $s$-$h$ phase transformation events happening together with, and assisted by, dislocational effects in the lattice, as some cells' strains respectively reach the $\rho$-well (hexagonal) or the $\zeta$-well (fully sheared square by a principal lattice-invariant shear) when the driving forces them away from $i$. We see  how, in the present variational GL-modelling, plastification may arise in the lattice via defect nucleation through lattice-invariant shears \cite{BCZZnature, ContiZanzotto, PRLgruppone, pacoreview, pacoreview2, pacoreview3,pacoreview4}, because in the reconstructive case the barriers to the these shears are only as high as the barriers relative to the phase transformation itself.

\begin{figure}

\centering
\includegraphics[width=7cm]{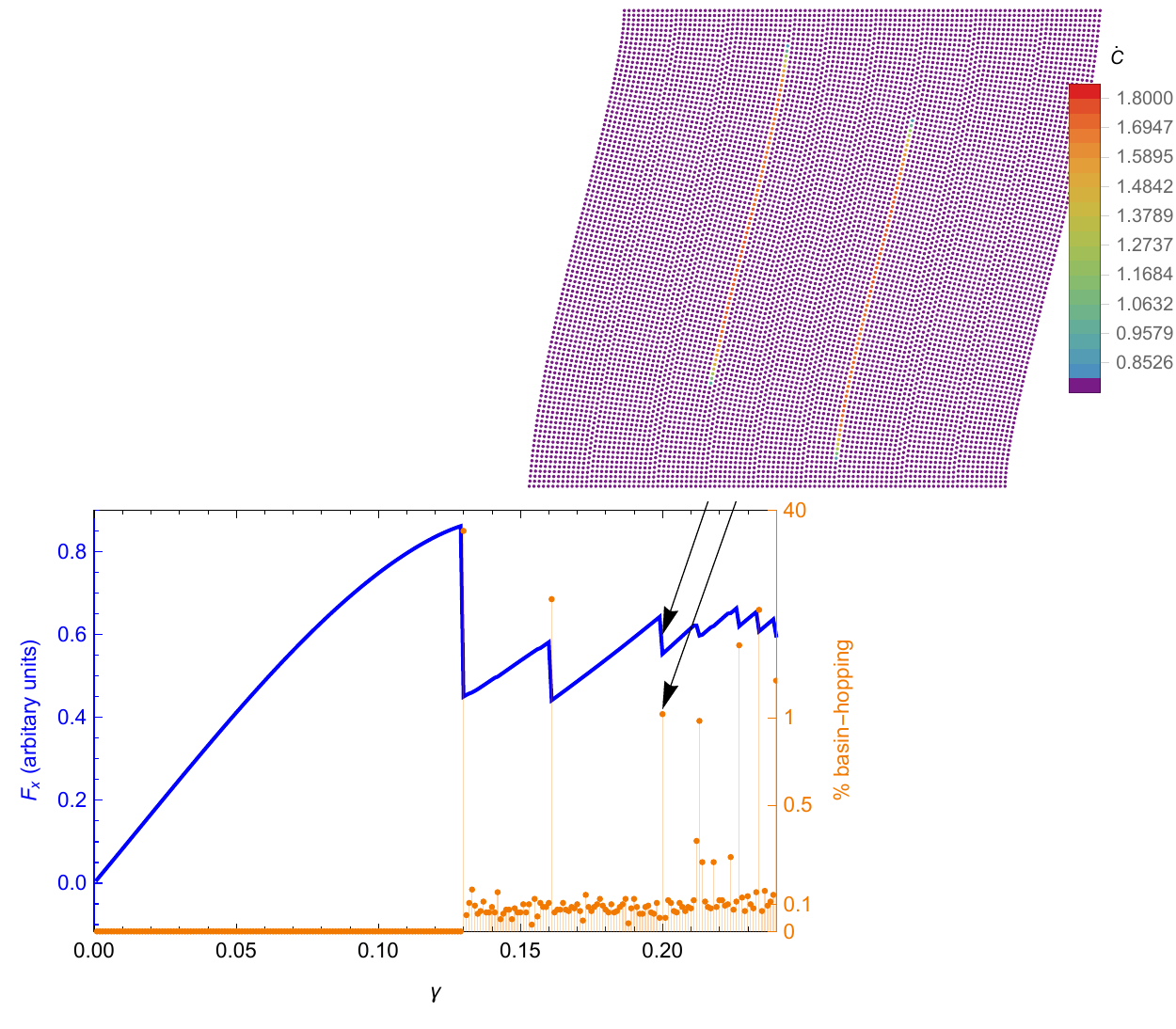}

\caption{\small
(Color online) Highlight of a strain burst characterizing the shear-driven $s$-$h$ transformation in Fig.~\ref{fig:snapshots}. This strain-evolution event in the square body undergoing quasi-static shear ($\gamma = 0.20$) corresponds to a spike (in orange) in the intermittent basin-hopping activity of the local strains on $\mathbb{H}$, with the associated relaxation drop in the stress-strain relation (blue jagged line). The strain avalanche is computed by considering the norm of the strain difference at each cell for two consecutive values of $\gamma$ in the simulation, near $\gamma = 0.20$. See Video V4 \cite{video4} in the SM for more details.
}
\label{cdot}
\end{figure}

Fig. \ref{nubecanyon} and Video V3 \citep{video3} in the SM display explicitly the shearing body's strain cloud as it flows in quasi-static intermittent fashion along the energy-surface valley floors. Most of the cell's strains are located near the involved well bottoms, with a fraction elastically stabilized on the non-convex regions between wells. Our simulations confirm the role of the GL-network of valley floors as giving the deformation pathways for the strain cloud on $\mathbb{H}$ during total-energy minimization, and show, as in \cite{IJP}, that these features of the energy landscape can help to better inform also other crystallographically-based approaches to crystal micromechanics, such as the phase-field models in \cite{biurzaza, corridoifrancesi}.

To conclude, we highlight explicitly the strain avalanches characterizing the intermittent deformation in this shear test, showing one such event in Fig.~\ref{cdot}. It originates from a burst of coordinated basin-hopping activity of the strain-values in $\mathbb{H}$, associated to microstructure and defect evolution in the lattice, producing a stress-relaxation event. The computed sequence of the strain avalanches observed in our quasi-static simulation is shown in Video V4 \cite{video4} of the SM. These results from the model qualitatively agree well with the strain-intermittency behavior experimentally observed during mechanically induced martensitic transformation in slowly driven shape-memory alloys, as evidenced in \citep{PRBbarrera}.

\bigskip
\noindent
{\it Aknowledgements}. We acknowledge the financial support of the Italian PRIN projects 2017KL4EF3,  2020F3NCPX\_001, and of INdAM-GNFM.

\section{Appendix: Online Supplementary Material}

\subsection{Caption to Supplementary Video V1 \cite{video1}}

Animation showing the GL-periodic $\beta$-bifurcation diagram for the critical points of the $s$-$h$ potential  $\sigma_\text{R}$ in \eqref{eq:rec}.

A section of this diagram is shown in Fig.~\ref{bifurcationvalleyfloors}(a). The Poincar\'e disk model \citep{michel, ContiZanzotto, tesipatriarca, bethesymmetree} is used here for the 2D hyperbolic space. The purple square and green circles in the disk respectively correspond to square and hexagonal points, with fat [slim] rhombic points given by bold [thin] blue lines, while red curves represent rectangular points. The 3D-diagram corresponds to increasing $\beta$ from bottom to top. Vertical lines indicate stable (bold) or unstable (dotted) square and hexagonal critical points. Stability ranges for these $s$-$h$ extremals are as detailed in Fig.~\ref{bifurcationvalleyfloors}(a) and in the text. In particular, there is the $s$-$h$ coexistence interval $\frac12<\beta<\frac32$ where both the square and hexagonal points are local minimizers, with branches of rhombic saddle points bifurcating from these two maximally symmetric ones with features and multiplicities as described in Fig.~\ref{bifurcationvalleyfloors}(a): three transverse rhombic branches issue from each hexagonal bifurcation point, while two rhombic branches issue with a subcritical pitchfork from each square bifurcation point. For completeness here are also indicated the supercritical pitchforks from square to rectangular local minimizers of the potential  $\sigma_\text{R}$ in \eqref{eq:rec} for $\beta \leq - \frac32$.

\subsection{Caption to Supplementary Video V2 \cite{video2}}

Shearing of a homogeneous square body containing an initially defect-free square lattice for an $s$-$h$ phase-transforming crystal with strain energy $\sigma_\text{R}$ in \eqref{eq:rec} with $\beta =1$. The imposed loading is along a primary shear direction in the square lattice, aligned with both the parallel square-cell side and bottom body side. The shearing boundary condition is imposed through the constrained horizontal sides of the body, with the two remaining sides free. See also Figs.~2-3-4 in the main text.

{\bf (a)} Bursty deformation field in the body for increasing shear parameter $\gamma$, indicated by the moving green dot along the $\gamma$-axis in {\bf (e)}. Lattice points are color-coded according to {\bf (c)} depending on the energy basin in the Poincar\'e half-plane $\mathbb{H}$ visited by the strain of each lattice cell during loading. Defect nucleation and evolution accompany the phase transformation, intrinsically produced in this model through energy minimization. See also Fig.~\ref{fig:snapshots}(d)-(e)-(f).

{\bf (b)}-{\bf (d)}-{\bf (f)} Intermittent evolution of the $\gamma$-dependent histograms of the four 2D deformation-gradient parameters recorded during the shear simulation.

{\bf (b)} Evolution of the 2D histogram (strain cloud) of the density of strain parameters on the Dedekind tessellation of $\mathbb{H}$ during shearing. The histogram color-coding provides the percentage of body cells with strain located at each point of $\mathbb{H}$. The straight horizontal dashed-blue line between the two neighboring square configurations $i$ and $i+1$ is the image in $\mathbb{H}$ of the primary shear path imposed as boundary condition. The initial configuration is in $i$ for $\gamma = 0$, while $\gamma = 1 $ corresponds to $i+1$. Shading indicates the convexity domains around the $s$-$h$ minimizers in $\mathbb{H}$ of the energy $\sigma_\text{R}$, while the red dashed lines mark the valley floor segments as in Fig.~3, which largely direct the strain-cloud evolution under the slow driving. See also Fig.~\ref{fig:snapshots}(a)-(b)-(c) and Fig.~\ref{nubecanyon}.

{\bf (c)} Color-code map used in {\bf (a)} for the GL-energy basins of $\sigma_\text{R}$ on the Dedekind tessellation of $\mathbb{H}$. The ridges marking the basins' boundaries are computed via a Eq.~\eqref{explicit valleyfloors} in the main text.

{\bf (d)} Evolution of the histogram for the values of $\det \bF$, indicating volumetric effects in the lattice.

{\bf (e)} Bursty $s$-$h$ phase transformation in the shearing test. Jagged stress-strain body behavior (blue), with the underlying spikes (orange) showing the percentage of basin-hopping strain values during loading, as $\gamma$ grows. The body response is elastic to about  $\gamma = 0.13$, where, after a first large stress drop,  intermittent stress-relaxation continues for growing $\gamma$. See also Fig.~\ref{fig:snapshots}(g) and Fig.~\ref{cdot}.

{\bf (f)} Evolution of the histogram for the values of the angle $\theta$ in the polar decomposition of the deformation gradient $\bF$, indicating local lattice rotation accompanying the phase transformation process. Notice the different rotation angles of the $s$-$h$ phase-microstructure bands, resulting in an evolving bimodal distribution for $\theta$.

\subsection{Caption to Supplementary Video V3 \cite{video3}}

Quasi-static bursty evolution of the strain field on $\mathbb{H}$ for growing parameter $\gamma$, shown directly on the energy landscape during the shear-driven $s$-$h$ transformation.  The GL-energy is $\sigma_\text{R}$ in Eq.~(\ref{eq:rec}) with $\beta = 1$, so the $s$-$h$ wells $i$, $\rho$, $\zeta$, $i+1$, ..., all have the same depth. Under the slow driving the strain values largely follow the energy valley floors indicated in dashed red in Fig.~\ref{bifurcationvalleyfloors}(b) and  Fig.~\ref{nubecanyon}. See also panel (b) in Video V2 \cite{video2} in the SM.

\subsection{Caption to Supplementary Video V4 \cite{video4}}

Strain avalanching during the shear-driven $s$-$h$ phase transformation.

{\bf (c)}-{\bf (a)} For reference, these two panels report respectively the evolution of the strain field in the shearing body, as in Video V2{\bf (a)}, and the associated jagged stress-strain relation, from Video V2{\bf (e)}.

{\bf (b)} Intermittent strain avalanching within the shearing crystal. Each event corresponds to phase transformation and defect evolution in the deforming lattice. Avalan\-ches are computed by considering the difference in strain norm at each lattice cell for two consecutive values of $\gamma$ in the simulation. The GL-energy is $\sigma_\text{R}$ in Eq.~(\ref{eq:rec}) with $\beta = 1$, as in Fig.~2. See also Fig.~6.

\end{document}